\documentclass[conference]{IEEEtran}
\IEEEoverridecommandlockouts
\usepackage{cite}
\usepackage{amsmath,amssymb,amsfonts}
\usepackage{algorithmic}
\usepackage{graphicx}
\usepackage{textcomp}
\usepackage{xcolor}
\usepackage[ruled,vlined]{algorithm2e}
\def\BibTeX{{\rm B\kern-.05em{\sc i\kern-.025em b}\kern-.08em
    T\kern-.1667em\lower.7ex\hbox{E}\kern-.125emX}}
\usepackage{adjustbox}
\usepackage{tabularx}
\usepackage{geometry}
\usepackage{xurl} 

\begin{document}

\title{Reinforcement Learning Based Penetration Testing of a Microgrid Control Algorithm}

\author{
\IEEEauthorblockN{Christopher Neal\IEEEauthorrefmark{1},
Hanane Dagdougui\IEEEauthorrefmark{2},
Andrea Lodi\IEEEauthorrefmark{2}, and
Jos\'e Fernandez\IEEEauthorrefmark{1}
}
\vspace{8pt}
\IEEEauthorblockA{\IEEEauthorrefmark{1}\textit{Department of Computer and Software Engineering} \\
\IEEEauthorblockA{\IEEEauthorrefmark{2}\textit{Department of Mathematics and Industrial Engineering}\\
\textit{Polytechnique Montreal}, Montreal, Canada\\
\{christopher.neal, hanane.dagdougui, andrea.lodi, jose.fernandez\}@polymtl.ca}
}
}

\maketitle

\begin{abstract}
Microgrids (MGs) are small-scale power systems which interconnect distributed energy resources and loads within clearly defined regions. However, the digital infrastructure used in an MG to relay sensory information and perform control commands can potentially be compromised due to a cyberattack from a capable adversary.  An MG operator is interested in knowing the inherent vulnerabilities in their system and should regularly perform Penetration Testing (PT) activities to prepare for such an event. PT generally involves looking for defensive coverage blindspots in software and hardware infrastructure, however the logic in control algorithms which act upon sensory information should also be considered in PT activities. This paper demonstrates a case study of PT for an MG control algorithm by using Reinforcement Learning (RL) to uncover malicious input which compromises the effectiveness of the controller. Through trial-and-error episodic interactions with a simulated MG, we train an RL agent to find malicious input which reduces the effectiveness of the MG controller.
\end{abstract}


\section{Introduction}
Penetration Testing (PT) is the process of performing an authorized attack on a system in order to uncover its vulnerabilities or to witness how the system will act in the presence of an attack \cite{pentest}. This is an accepted activity which is practiced by specialized professionals, studied by academia, and has become an established industry for assessing the defensive posture of a system. Automated PT tools, such as MetaSploit, can aid in the process, however current tools rely upon handcrafted attack rules developed by a programmer and guided by a domain expert \cite{metasploit_book}. This paper proposes the use of a form of Artificial Intelligence (AI) known as Reinforcement Learning (RL) as a methodology to automate the  generation of PT attacks and evaluates this approach using attacks on a simulated Microgrid (MG).

MGs are small-scale electrical distribution networks which can generate, store, and supply local consumers through the use of Distributed Energy Resources (DERs) (e.g. photovoltaic solar panels, wind turbines, diesel generators) and Energy Storage Systems (ESSs) (e.g. batteries) \cite{opf_mg, trends_mg}. MGs may act autonomously in island-mode or they can operate in grid-connected mode where they can exchange electricity with the main electrical grid, with the capacity to realize a transition between the two modes. The adoption of MGs introduces a number of cybersecurity concerns, since the operation of an MG is generally performed by digital devices over wireless networks~\cite{mg_cyber}. MGs are an example of a Cyber-Physical System (CPS), where there are physical processes controlled by cyber/digital-based software \cite{CPS_survey}. This type of environment provides a useful arena for witnessing the impact of cyberattacks, since the effect of an attack can be directly measured by the physical processes in operation. This paper considers training an RL agent to craft False Data Injection (FDI) attacks which modify the input values provided to a Microgrid Central Controller (MGCC) operating in a simulated MG. The methodology utilized here to find vulnerabilities in an MGCC could be extended to networked MGs or other Industrial Control System (ICS) environments.




\section{Background and Related Work}\label{sec_background}
PT is one of several activities in establishing a defensive posture of a computing-reliant system, such as an MG. An initial defense in securing an MG involves adhering to best practices outlined by industry norms. The National Institute of Standards and Technology (NIST) has outlined standard cybersecurity vulnerability classes for Smart Grids and MGs~\cite{NIST}. The Sandia National Laboratories encourages the use of a defense-in-depth MG infrastructure architecture in preparation for potential attacks \cite{Sandia_MG}. The Mitre Corporation has expanded its ATT\&CK framework to consider typical attack tactics and techniques in ICSs, such as MGs \cite{MITRE_ICS, MITRE_PRESS}. 

Despite an organization's best efforts, various attack vectors are likely to exist through vulnerable components in MG installations, including Programmable Logic Controllers (PLCs), Smart Meters, Phasor Measurement Units (PMUs), and Phasor Data Concentrators (PDCs) \cite{7152015, Sandia_MG}. These components are susceptible to malware delivery, software misconfiguration,  Denial-of-Service (DoS), and eavesdropping over Transmission Control Protocol and Internet Protocol (TCP/IP) networks. Compromised reporting devices can be subjugated to FDI attacks, which involves sending falsified values amongst MG devices with malicious intent. FDI attacks have been studied for MGs \cite{8953044,7301476,chlela_attack}, along with defensive measures \cite{8550549, 8051297,7829372}. Although precautions should be taken in securing reporting devices in an MG, it is wise to consider the scenario where devices may be compromised and report falsified data to control algorithms. Experimentation with MG attacks and defenses using testbeds can be used to this end without causing any negative real-world effects \cite{HOSSAIN2014132,6392520}. 

This paper considers training an RL agent to learn an effective attack strategy in a simulated MG  environment. Successes in RL-based game-playing agents for Go \cite{goarticle} and StarCraft~II \cite{starcraft} have shown the ability of RL agents to uncover strategy-spaces not considered by the best human players, which also may be the case for PT activities. 

RL can be characterized as a form of statistical learning where an agent learns to perform actions in an environment with the goal of maximizing some notion of long-term reward\cite{sutton}. The agent's goal is to learn a near-optimal policy $\pi$, through repeated trial-and-error episodic interactions with an environment. There is a balance to be played in exploring unseen decision-spaces while exploiting known `good' decision-spaces. RL problems are often modelled as some form of a Markov Decision Process (MDP), where there exists a set of states $S$, a set of actions $A$, the probability $P_a(s,s')$ to transition from state $s$ to state $s'$ given action $a$, and the reward $R_a(s,s')$ for transitioning from state $s$ to $s'$ using action $a$. Formulating PT activities as an RL problem involves defining the testbed environment's state, the set of actions an attacking agent can perform, the reward that the agent receives for taking actions, and a method for updating the agent's policy. Using RL for PT has been previously proposed for traditional Information Technology (IT) environments  \cite{8611595,pomdp_hackers,mitre_mdp}, however this is the first attempt, to our knowledge, to use RL for PT of an MG control algorithm.

\section{Reference Microgrid Testbed}\label{sec_microgrid}

\subsection{Microgrid Architecture}
\begin{figure}[h!]
\includegraphics[width=\linewidth]{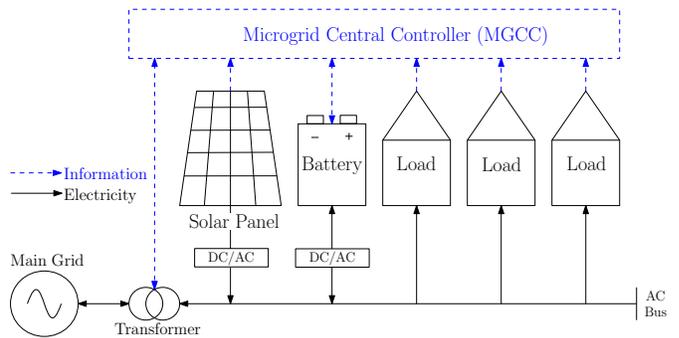}
\caption{Microgrid Architecture}
\label{fig:mg_architecture}
\end{figure}

This paper considers a small residential MG consisting of a Photovoltaic (PV) system (i.e. solar panel), a Battery Energy Storage System (BESS), and three households acting as fluctuating loads over time (Fig.~\ref{fig:mg_architecture}). The solar panel generates electricity relative to the sun's solar irradiance and is used to supply electricity to the households. Excess electricity generated by the solar panel can either be sold to the grid operator or stored in the BESS to be used at a later time. Note that the MG is connected to the main electrical grid via a grid-connected transformer. Should the MG not be able to satisfy the load demand, electricity can be bought from the grid operator. The MGCC receives sensory information about the given state of the MG and determines control actions to satisfy the electrical load while minimizing total cost.

This paper uses an MG configuration that comes with MATLAB/Simulink as the simulation environment \cite{MATLAB}. This model has been considerably modified to enable the utilization of custom controller logic and the ability to insert falsified values from simulated attacks. Controlling this MG involves switching the battery into an \texttt{ON} or \texttt{OFF} state. There are four control scenarios which are dictated by how much power is being generated by the solar panel (see Table~\ref{table:battery_control}).

\begingroup
\renewcommand{\arraystretch}{1.25} 

\begin{table}[h!]
\caption{Battery control scenarios}
\begin{center}
\begin{tabular}{|c||c|c|}
\hline
	& \small{\textbf{Battery \texttt{ON}}} & \small{\textbf{Battery \texttt{OFF}}} \\
\hline	
\small{\textbf{Solar Power $\geq$ Load}} & \small{Charge battery} & \small{Sell to grid} \\
\hline	
\small{\textbf{Solar Power $<$ Load}} & \small{Discharge battery} & \small{Buy from grid} \\
\hline
\end{tabular}
\label{table:battery_control}
\end{center}
\end{table}
\endgroup
By default, any generated solar power is used to satisfy the load. Any excess solar power can be utilized to charge the battery or can be sold to the main grid. When there is a deficit between the solar power and the load, this missing quantity can be met from the battery reserve or can be bought from the grid. Note that in reality, other battery control scenarios would exist. For example, a combination of grid and battery power could be used to satisfy the load, or, any excess solar power could be split to have a portion charge the battery and have the other portion sold to the grid. These scenarios are not considered to favor a more tractable simulation setting.

Simulation scenarios are run for this MG by loading in comma-separated-values (CSV) files for the solar irradiance, household loads, electricity selling price, and electricity buying price. 

\subsection{Optimization Based Controller}\label{controller}
\begin{figure}[h!]
\includegraphics[width=\linewidth]{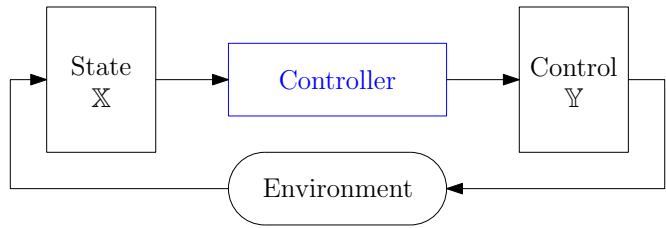}
\caption{Control Loop Overview}
\label{fig:control_loop}
\end{figure}

The controller is tasked with satisfying the load demand while minimizing the total cost. Purchasing power from the grid increases the total cost and selling excess power reduces the total cost. No preexisting controller software could be readily used in this MG simulation, hence a controller based on  mathematical optimization has been developed using CPLEX and the OPL language \cite{CPLEX_REF}. At each hour in the simulation, the controller is provided an input set describing the state $\mathbb{X}$ and determines the control set $\mathbb{Y}$ to drive the MG in a cost-effective manner. This process is outlined in Fig.~\ref{fig:control_loop}. 

At each timestep the controller is provided the current state values as well as forecasted values over a time window of length $T=24$. The controller determines the optimal control actions over the entire time horizon and then sends the control signal to the MG for the current timestep.

The state is initially defined as $\mathbb{X} = \{ P^{L},P^{P},\Lambda^{B},\Lambda^{S},b \}$. The set $P^{L} = \{p_{1}^{L}, p_{2}^{L}, ..., p_{T}^{L}\}$ is the expected cumulative load power demand values from the households over the window, where $p_{1}^{L}$ is the initial load power and $p_{2}^{L},...,p_{T}^{L}$ are the forecasted loads. The set $P^{P} = \{p_{1}^{P}, p_{2}^{P}, ..., p_{T}^{P}\}$ is the expected generated photovoltaic solar power values over the control horizon, where $p_{1}^{P}$ is the initial solar power and $p_{2}^{P},...,p_{T}^{P}$ are the forecasted values. The buying and selling prices over the window are $\Lambda^{B} = \{\lambda_{1}^{B}, \lambda_{2}^{B}, ..., \lambda_{T}^{B}\}$ and $\Lambda^{S} = \{\lambda_{1}^{S}, \lambda_{2}^{S}, ..., \lambda_{T}^{S}\}$, respectively. The initial battery state of charge is $b$.

To simplify the control problem, the basic formulation is updated by using $P^L$, $P^P$, $\Lambda^{B}$, and $\Lambda^{S}$ to precalculate two new sets, Costs $C$ and Differences $D$. These sets are calculated using Algorithm~\ref{costdiff}.

In the algorithm, the value $e_t$ is a temporary variable to hold the difference between the solar power and load demand at time $t$. The value $c_t$ indicates how the overall cost of the controller is affected at time $t$. If $c_t < 0$ there is a surplus of solar power that can be sold to the grid and if $c_t > 0$ there is a shortfall of solar power which can be bought from the grid. The value $d_t$ indicates how the state of charge of the battery is affected at time $t$. The constant $\omega$  represents the maximum power quantity of the battery and is used to convert the battery power to a percentage, representing the battery's state of charge. If $d_t < 0$ there is shortfall of solar power which can be retrieved from the battery and if $d_t > 0$ there is an excess of solar power which can be supplied to the battery. The optimization problem will determine at each time $t$ to either update the controller's total cost by taking quantity $c_t$ or update the battery's state charge by taking quantity $d_t$. The updated version of the state is  $\mathbb{X} = \{ C,D,b \}$

\begin{algorithm}[h!]
\SetAlgoLined
$C := \emptyset$ \\
$D := \emptyset$ \\
 \For{$t=1$ to $T$}{
	$e_t := p_{t}^{P} - p_{t}^{L}$ \\
  \eIf{$e_t \geq 0$}{
   $c_{t} := - \lambda_{t}^{S}e_t$ \\
   }{
   $c_{t} := - \lambda_{t}^{B}e_t$ \\
   }
   $d_t = e_t / \omega$ \\
   $C := C \cup c_t$ \\
   $D := D \cup d_t$ \\
 }
 \Return $C$, $D$
 \caption{Construct Costs and Differences sets}
 \label{costdiff}
\end{algorithm}

Consider the example sets $P^P = \{160,140,100,0\}$, $P^L = \{100,120,170,250\}$, $\Lambda^{B}=\{3,3,2,2\}$, and $\Lambda^{S}=\{1,1,1,1\}$, with $\omega=10$. The algorithm will return $C=\{-60,-20,140,500\}$ and $D=\{6, 2, -7, -25\}$. 

The controller uses the state $\mathbb{X}$ as input to calculate the control set $\mathbb{Y}=\{Y, \bar{Y}, B\}$. The set $Y=\{y_1,y_2,...,y_T\}$ is the expected set of control commands over the control horizon, where $y_t = 1$ implies that the battery is expected to be in an \texttt{ON} state at time $t$ and $y_t = 0$ implies that the battery is expected to be in an \texttt{OFF} state at time $t$. The set $\bar{Y} = \{ \bar{y}_1, \bar{y}_2, ..., \bar{y}_T\}$ is a control variable which represents the opposite of the battery control and can be considered as the grid control command. If $y_t = 1$, then $\bar{y_t}=0$, and vice versa. This is required to solve the controller's optimization problem which is a Mixed Integer Linear Programming (MILP) problem.  The set $B = \{b_{1}, b_{2}, ..., b_{T}\}$ is the predicted battery state of charge over the window and is calculated to ascertain the battery capacity constraints are satisfied. After solving for $Y$, $ \bar{Y}$, and $B$, the command value $y_1$ is provided to the system as the invoked control in the simulation. The simulation will then proceed to execute until the next hour where the process is repeated, and so on.  

At each timestep, the controller is given $\mathbb{X}$ and solves Equation~\eqref{eq1:form} by choosing $\mathbb{Y}$ which minimizes the cost function $f$, subject to satisfying the set of operating constraints $g$.

\begin{subequations}\label{eq1:form}
\begin{align}
\text{min cost} \quad &= \quad \underset{\mathbb{Y}}{\text{min}} \quad f(\mathbb{X},\mathbb{Y)} \label{eq1:l1}\\
\quad & \quad \quad ~~ \text{s.t.} \quad g(\mathbb{X}, \mathbb{Y}) \leq 0. \label{eq1:l2}
\end{align}
\end{subequations}

The fully expanded optimization control problem is modelled by Equation~\eqref{eq2:formulation}:

\begin{subequations}
\label{eq2:formulation}
\begin{align}
    \underset{Y,\bar{Y} \in{\{0,1\}^T}, B\in{\mathbb{R}_{+}^{T}} }{\text{min}} 
    			& \quad \sum_{t=1}^{T} \bar{y_t}c_{t}   \label{eq2:l1}\\
    \text{s.t.} & \quad b_0 = b \label{eq2:l2} \\
    			& \quad \text{for}~t=1,...,T \nonumber \\
        		& \quad b_t = b_{t-1} + y_t d_{t} \Delta^{\tau}  \label{eq2:l3} \\
        		& \quad y_t + \bar{y_t} = 1 \label{eq2:l4}\\
        		& \quad b^{\text{min}} \leq b_t \leq b^{\text{max}}. \label{eq2:l5}
\end{align}
\end{subequations} 

Equation~\eqref{eq2:l1} states that the total controller's cost is the sum of the values $c_t$ when the battery is \texttt{OFF} (i.e. $\bar{y_t}=1$ and $y_t=0$). Equation~\eqref{eq2:l2} sets the initial battery state of charge. Equation~\eqref{eq2:l3} states that the battery state of charge at time $t$ is equal to the previous charge plus the value $d_t$ if the battery is \texttt{ON} (i.e. $y_t=1$ and $\bar{y_t}=0$), times the length of the step time $\Delta^{\tau}$. Equation~\eqref{eq2:l4} is used to ensure that the battery is either in an \texttt{ON} or \texttt{OFF} state. Lastly, Equation~\eqref{eq2:l5} keeps the battery state of charge within its allowed operating range, where $b^{\text{max}}$ is the battery's maximum state of charge and $b^{\text{min}}$ is the battery's minimum state of charge.

\subsection{Threat Model}
We assume the attacker to be a skilled adversary which has the capability to undermine the MG's digital-based infrastructure in such a way they can modify the controller's input $\mathbb{X}$. The attacker can insert a malicious set of input $\mathbb{A}^{\mathbb{X}}$, as shown in Equation~\eqref{eq3}, into the state values so that the controller receives the updated set $\mathbb{X}^{\mathbb{A}}$ as input:
\begin{equation}\label{eq3}
	\mathbb{X}^{\mathbb{A}} = \mathbb{X} + \mathbb{A}^{\mathbb{X}}.
\end{equation}

This paper specifically considers the case where the attacker modifies the initial battery state of charge that is reported to the controller. The attacker's problem of choosing malicious input can be represented as a Bilevel Programming Problem (BPP), where the attacker constructs an input which subverts the goal of the controller and intends to maximize the total cost of the controller. Consider the attack shown in Equation~\eqref{eq6:formulation} where the attacker determines some value $a^b$ to add to the initial battery state of charge $b$, in order to create an attacked initial battery state of charge $b^A$, which will be used by the controller in its control problem.

\begin{subequations}
\label{eq6:formulation}
\begin{align}
\underset{a^{b}\in{\mathbb{R}}, b^{A}\in{\mathbb{R}_{+}}}{\text{max}} & \quad \sum_{t=1}^{T} \bar{y_t}c_{t}  \label{eq6:l1}\\
								 \text{s.t.} & \quad b^{A} = b + a^{b} \label{eq6:l2}\\
								             & \quad a^{b^{\text{min}}} \leq a^{b} \leq a^{b^{\text{max}}} \label{eq6:l3}\\
								             & \quad b^{\text{min}} \leq b^A \leq b^{\text{max}} \label{eq6:l4}\\	
								             & \underset{Y,\bar{Y} \in{\{0,1\}^T}, B\in{\mathbb{R}_{+}^{T}} }{\text{min}} 
    			 \quad \sum_{t=1}^{T} \bar{y_t}c_{t} \label{eq6:l5} \\	
    			 						     & ~~~~~~~~~~~~~~~~~~~\text{s.t.} \quad b_0 = b^A \label{eq6:l6} \\         
    			 						     & ~~~~~~~~~~~~~~~~~~~~~~~~~ \text{for}~t=1,...,T \nonumber \\
    			 						     & ~~~~~~~~~~~~~~~~~~~~~~~~~ b_t = b_{t-1} + y_t d_{t} \Delta^{\tau}  \label{eq6:l7} \\
    			 						     & ~~~~~~~~~~~~~~~~~~~~~~~~~ y_t + \bar{y_t} = 1 \label{eq6:l8}\\
    			 						     & ~~~~~~~~~~~~~~~~~~~~~~~~~ b^{\text{min}} \leq b_t \leq b^{\text{max}}. \label{eq6:l9}
\end{align}
\end{subequations} 

BPPs are hard to solve to optimality since they are generally non-convex, non-differentiable, and have been shown to be at least NP-hard \cite{jeroslow}. However, it is in the interest of an MG operator to understand which malicious values could be inserted into their system and have it operate in a sub-optimal manner. Due to the difficulty of solving such BPPs, this paper proposes the use of RL as a metaheuristic to uncover approximate optimal solutions to this MG attack problem.

\section{Evaluation}\label{sec_evaluation}

\subsection{Experimental Setup}\label{sub_setup}

\begin{figure*}[h!]
\centering
\includegraphics[width=\linewidth]{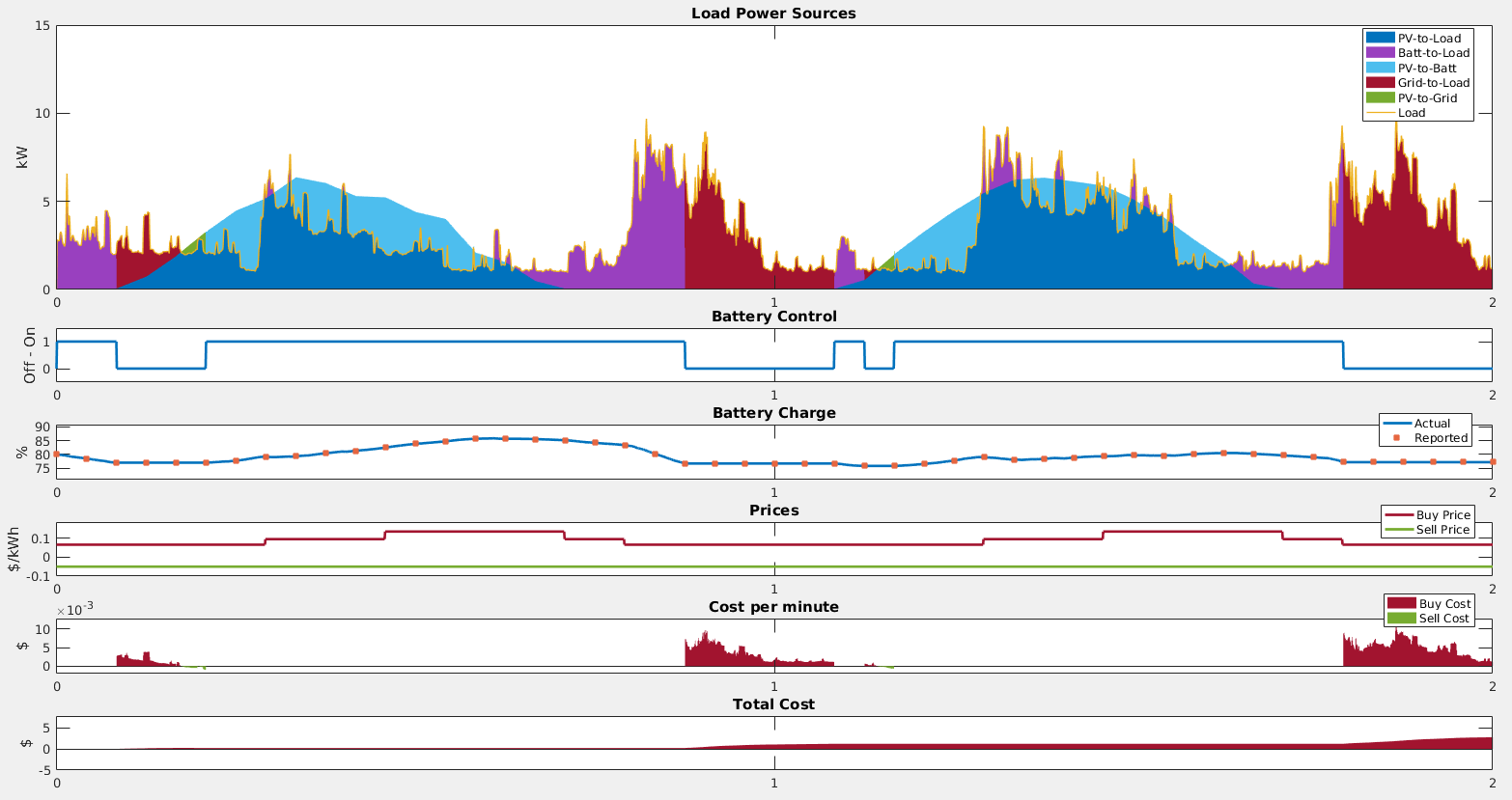}
\caption{Scenario: No attack. Score: \$2.74. Avg Batt Charge: 79.301\%}
\label{fig:controller_mincost}
\end{figure*}

The MG simulation uses CSV files as input parameters of household loads, solar irradiance, electricity purchasing prices, and electricity selling prices. All the experiments use identical input from an example period of 48 hours. The household loads come from a freely accessible dataset of one minute samples of household power consumption from a household in Sceaux, France over a four year period \cite{uciml}. The three MG households are represented by the days of October 6-7, 2008, October 12-13, 2009, and October 11-12, 2010. These represent typical 48-hour electricity usage and summed together are considered as the load of the MG. Solar irradiance data could not be found from Sceaux, France, so solar data was taken from a sunny day in Ottawa, Canada for the days of July 20-21, 2015 \cite{weatherstats}. Although, the data sources do not come from the same location, taken together they form an interesting scenario and demonstrate the abilities of the  controller.

As the simulation runs, the controller invokes a command at each hour. The cost accrued over the simulation is calculated for each minute, based on whether electricity was bought or sold over the minute. Buying electricity increases the cost, while selling electricity decreases the cost. The buying price varies throughout the day at fixed intervals (similarly to some real-world consumer electrical prices \cite{ontariohydro}):
\begin{itemize}
\item \textbf{Off-peak:} 6.5 cents/kWh (19:00-07:00)
\item \textbf{Mid-peak:} 9.4 cents/kWh (07:00-12:00, 17:00-19:00)
\item \textbf{On-peak:} 13.4 cents/kWh (12:00-17:00)
\end{itemize}
Any excess power can be sold to the grid at a fixed price.
\begin{itemize}
\item \textbf{Constant:} 5.0 cents/kWh
\end{itemize}

At each hour in the simulation, the controller receives the current loads, solar irradiance, and prices, as well as their forecasts for the next 24 hours. The forecasts are 100\% accurate in these experiments and the controller has perfect information to work with. The controller is also given the current reported battery state of charge. The controller solves Equation~\eqref{eq2:formulation} each hour to determine the command to invoke. The battery is restricted to operating between 75\%-100\% of its capacity. This prevents the controller from utilizing all available power and not maintaining any reserves for an emergency. 

When the controller runs in normal operating conditions without any attack present, we get the results shown in Fig.~\ref{fig:controller_mincost}. The reported battery state of charge match with the actual battery charge. The controller algorithm reduces the overall operating cost by performing \textit{peak shaving} which utilizes the battery power to offset peak load demand in the evening.

\subsection{Attack Training Results}\label{sub_learning}
This paper's threat model assumes that an attacker has the ability to modify the battery state of charge $b$, which is passed to the controller algorithm. The RL learning agent is tasked in determining what is the value to provide to the controller as input, causing the controller to make sub-optimal decisions and increase the overall cost incurred. We utilize an implementation of the Temporal Difference (TD) Advantage Actor-Critic algorithm \cite{ac_algo}, since the RL agent's action involves choosing a continuous value \cite{sutton_policy, 6392457}.

An episode consists of a two-day period with $K=48$ total hourly timesteps where the RL agent makes a decision. For each hourly timestep $k=1,...,K$, the RL agent receives the state $s_k=\{b_k,c_k,d_k,p^L_k,p^P_k\}$, where $b_k$ is the current battery state of charge, $c_k$ is the cost value over the current hour, $d_k$ is the difference value over the current hour, $p^L_k$ is the current load power, and $p^P_k$ is the current solar power. The algorithm samples an action from the current policy to determine the value $a^b$ to add to $b$. The input is provided to the controller and the simulation runs until the next hourly timestep. The total cost of the controller over the hour is summed (and multiplied by 1000), then returned as the reward $r_k$ along with the new state of the system $s'_k$. The actor-critic networks are updated to minimize the actor and critic total loss, with the goal of finding optimal weight parameters for maximizing the total reward. 

\begin{figure}[h!]
\centering
\includegraphics[width=\linewidth]{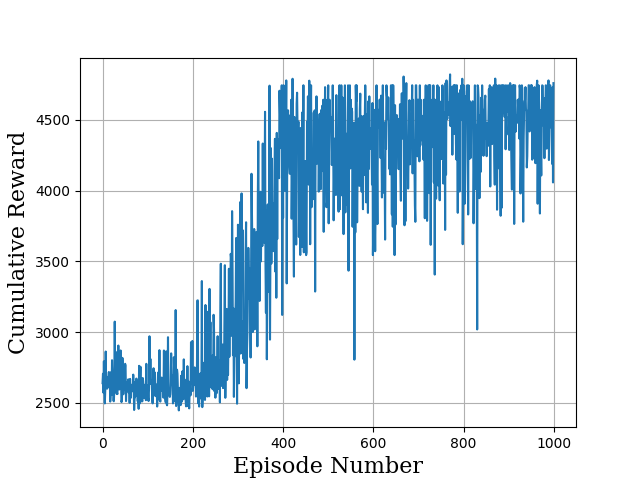}
\caption{Training of RL agent which can choose any value for the battery charge}
\label{fig:train_ANY}
\end{figure}

Fig.~\ref{fig:train_ANY} shows the cumulative reward of the trained RL agent over 1000 episodes. This agent is permitted to choose any value for $a^b$, while respecting the requirement that the state of charge must remain between 75\% and 100\%. The cumulative reward trends upward as the agent learns an effective strategy for this scenario, converging near a value of 4600.

For comparison, another agent was trained which only has the ability to modify the reported battery state of charge by plus or minus 5 percent of the actual battery state of charge. These training results are not shown, however the capability of this agent in increasing the total cost is more limited and there is a more gradual convergence.

\subsection{Attack Analysis}\label{sub_attack}
\begin{figure*}[h!]
\centering
\includegraphics[width=\linewidth]{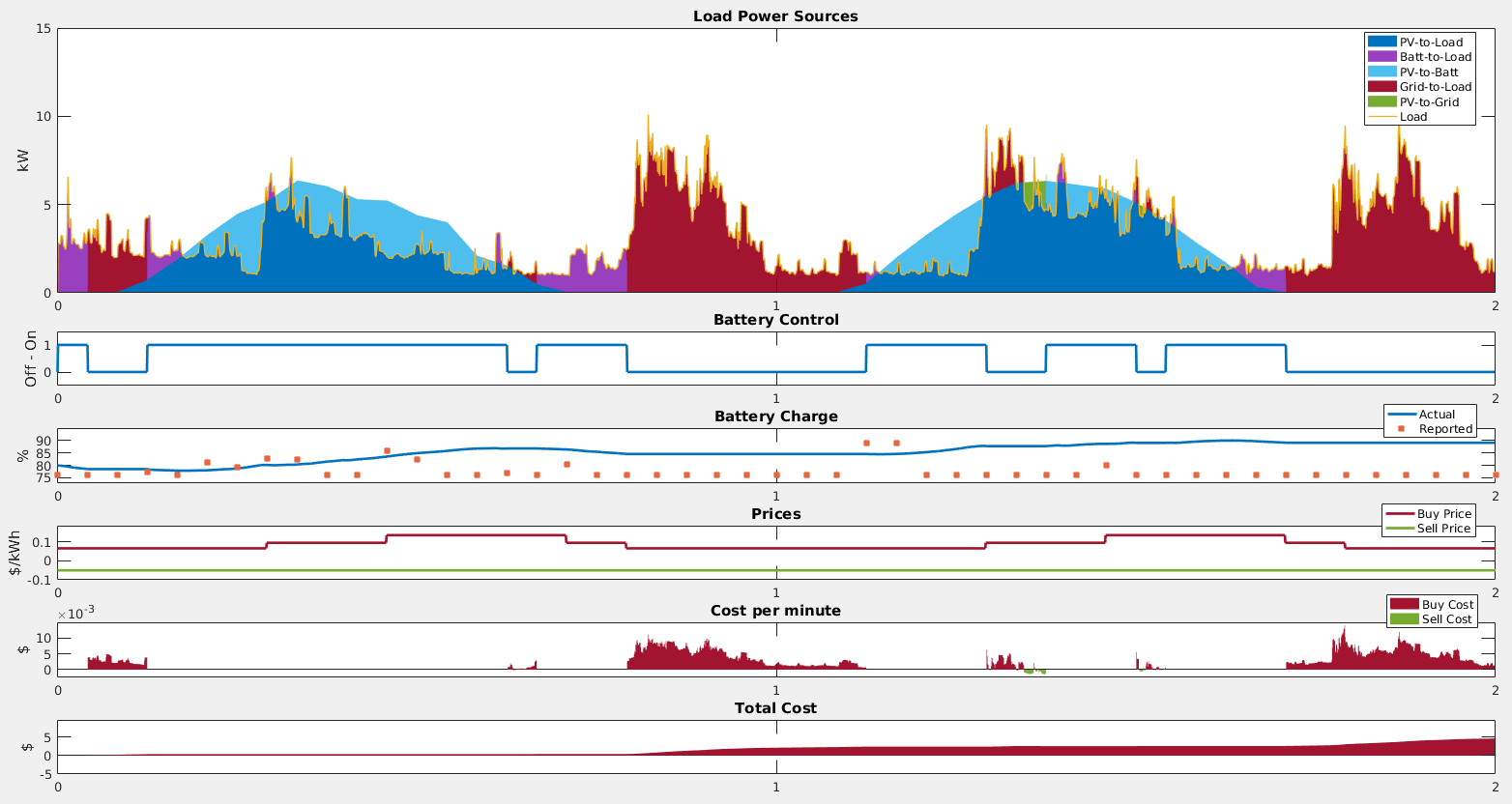}
\caption{Scenario: Attack reported battery charge ($75\text{\%}\leq b^A \leq 100\text{\%}$). Score: \$4.55. Avg Batt Charge: 85.262\%}
\label{fig:controller_rlattack_any}
\end{figure*}

The trained RL agents from Section~\ref{sub_learning} are used to attack the reported battery state of charge level sent to the controller algorithm. Fig.~\ref{fig:controller_rlattack_any} shows a scenario with an initial battery state of charge of 80\% and where the RL agent can choose any value for $a^b$ such that $75\text{\%}\leq b^A \leq 100\text{\%}$. The attacker generally reports that the battery state of charge is near the allowed lower bound of 75\%, causing the controller to believe there is little reserves to be used to satisfy the load. The effect is that the controller is less effective at performing \textit{peak shaving}, causing the battery to not offset the demand, and the state of charge of the battery to rise. The overall score is substantially higher than the same scenario in Fig.~\ref{fig:controller_mincost} where there is no attack. 

A more detailed analysis of the behavior of the attacking agents is provided in Table~\ref{table:attack_results}. Attacks are carried out in the simulation over the 48-hour period with different initial battery state of charge levels. Each scenario is run 10 times and the values are averaged. In general, the attacking agents report a battery state of charge as low as possible. This causes the controller to not use its reserve battery power to supply the load and hence the overall battery charge increases. 
	
	\begingroup
	\renewcommand{\arraystretch}{1.15} 
	\begin{table*}[h!]
		\caption{Simulation results for 48-hour period with varying initial battery charge values and attack scenarios}
	\begin{adjustbox}{width=\textwidth}
	\centering
	\begin{tabular}{|c||c|c|c|c||c|c|c|c|}
	\hline
	 & \multicolumn{4}{c||}{\textbf{Init. Batt. Charge: 75\%}} & \multicolumn{4}{c|}{\textbf{Init. Batt. Charge: 80\%}} \\
	\hline
	\textit{Attack} & Cost (\$) & Cost Incr. & Avg Charge & Avg Reported &  Cost (\$) & Cost Incr. & Avg Charge & Avg Reported \\
	\hline
	\textit{None}           			& 3.62 & -        & 79.64\% & 79.62\% &	2.74  & -        & 79.30\% & 79.30\% \\
	\textit{$b^A+/- 5$\%}  				& 3.95 & +9.17\%  & 80.80\% & 79.44\% & 3.41  & +24.45\% & 82.21\% & 80.18\% \\
	\textit{$75 \%\leq b^A \leq 100 \%$}  & 4.13 & +14.22\% & 80.47\% & 78.99\% & 3.83  & +39.83\% & 83.08\% & 78.62\% \\
	\hline
	\hline
	 & \multicolumn{4}{c||}{\textbf{Init. Batt. Charge: 85\%}} & \multicolumn{4}{c|}{\textbf{Init. Batt. Charge: 90\%}} \\
	\hline
	\textit{Attack} & Cost (\$) & Cost Incr. & Avg Charge & Avg Reported &  Cost (\$) & Cost Incr. & Avg Charge & Avg Reported \\
	\hline
	\textit{None}           			& 2.48 & -         & 82.58\% & 82.58\% &	1.87  & -        & 83.94\% & 83.99\% \\
	\textit{$b^A +/- 5$\%} 				& 2.97 & +19.67\%  & 86.36\% & 82.22\% &    2.47  & +31.99\% & 88.24\% & 83.79\% \\
	\textit{$75 \%\leq b^A \leq 100 \%$}  & 4.41 & +77.76\%  & 89.80\% & 77.20\% &    4.60  & +146.47\% & 95.46\% & 76.39\% \\
	\hline
	\hline
	 & \multicolumn{4}{c||}{\textbf{Init. Batt. Charge: 95\%}} & \multicolumn{4}{c|}{\textbf{Init. Batt. Charge: 100\%}} \\
	\hline
	\textit{Attack} & Cost (\$) & Cost Incr. & Avg Charge & Avg Reported &  Cost (\$) & Cost Incr. & Avg Charge & Avg Reported \\
	\hline
	\textit{None}           			& 1.58 & -         & 86.07\%  & 86.14\% &	1.06  & -         & 87.30\%  & 87.39\% \\
	\textit{$b^A+/- 5$\%}  				& 1.74 & +10.63\%  & 88.29\%  & 84.28\% &   1.45  & +36.41\%  & 90.08\%  & 85.64\% \\
	\textit{$75 \%\leq b^A \leq 100 \%$}  & 4.65 & +194.97\% & 100.00\% & 76.10\% &   4.75  & +348.15\% & 100.00\% & 76.03\% \\
	\hline
	\end{tabular}
		
		\label{table:attack_results}
	\end{adjustbox}	
	\end{table*}
	\endgroup
	
\subsection{Discussion}\label{sub_discussion}	
The implemented MG testbed is simplistic by design and enables the goal of training RL agents to perform PT activities on a MG control algorithm. This paper demonstrates an approach of formalizing a control algorithm and then uses RL experimentation to understand its vulnerabilities. This serves as an introductory study of this approach, however more insights could be derived from a more realistic MG environment. Some areas to consider include having a greater number of MG components, arranging the MG in a radial lateral configuration, using networked MGs, and considering the effects of voltage regulation activities. Additionally, the described testbed is purely a simulation of a few MG components and their dynamics. More insights could be attained by simulating a communications network and even incorporating actual controller software or hardware.

The results from the experiments show that an attack which is able to report a lower battery state of charge than what it is in reality, will have the most impact on the effectiveness of the given controller algorithm. This result may seem intuitive, however the proposed experimental process  concretely verifies this intuition and would guide a potential defender in prioritizing resources. Performing such an activity in a more complex MG, or any ICS, environment may yield more profound and unidentified conclusions. 

\section{Conclusion}\label{sec_conclusion}
This paper presents a framework for performing PT activities against MG control algorithms using RL and has been applied to a simulated MG as a case study. The MG architecture is implemented in MATLAB/Simulink and is controlled by an algorithm formulated as an MILP which is implemented using CPLEX. The threat model identifies the reported battery state of charge as a threat vector and RL based agents are trained to learn how to modify this value to negatively affect the effectiveness of the controller. The generated attacks show that reducing the reported battery state of charge to the controller has the largest impact on its overall performance. This PT framework could be applied to networked MGs and other ICS environments to uncover inherent vulnerabilities in control algorithms.

\bibliographystyle{unsrt}
\bibliography{\jobname}

\begin{thebibliography}{10}

\bibitem{pentest}
Patrick Engebretson.
\newblock {\em {The Basics of Hacking and Penetration Testing: Ethical Hacking
  and Penetration Testing Made Easy}}.
\newblock Elsevier, 2nd edition, 2013.

\bibitem{metasploit_book}
David Kenney, Jim O'Gorman, Devon Kearns, and Mati Aharoni.
\newblock {\em {Metasploit: The Penetration Tester's Guide}}.
\newblock No Starch Press, 1st edition, 2011.

\bibitem{opf_mg}
E.~{Dall'Anese}, H.~{Zhu}, and G.~B. {Giannakis}.
\newblock Distributed optimal power flow for smart microgrids.
\newblock {\em IEEE Transactions on Smart Grid}, 4(3):1464--1475, 2013.

\bibitem{trends_mg}
D.~E. {Olivares}, A.~{Mehrizi-Sani}, A.~H. {Etemadi}, C.~A. {Cañizares},
  R.~{Iravani}, M.~{Kazerani}, A.~H. {Hajimiragha}, O.~{Gomis-Bellmunt},
  M.~{Saeedifard}, R.~{Palma-Behnke}, G.~A. {Jiménez-Estévez}, and N.~D.
  {Hatziargyriou}.
\newblock Trends in microgrid control.
\newblock {\em IEEE Transactions on Smart Grid}, 5(4):1905--1919, 2014.

\bibitem{mg_cyber}
Silvia Marzal, Robert Salas, Raúl González-Medina, Gabriel Garcerá, and
  Emilio Figueres.
\newblock Current challenges and future trends in the field of communication
  architectures for microgrids.
\newblock {\em Renewable and Sustainable Energy Reviews}, 82:3610 -- 3622,
  2018.

\bibitem{CPS_survey}
A.~{Humayed}, J.~{Lin}, F.~{Li}, and B.~{Luo}.
\newblock Cyber-physical systems security—a survey.
\newblock {\em IEEE Internet of Things Journal}, 4(6):1802--1831, 2017.

\bibitem{NIST}
{National Institute of Standards and Technology (NIST): U.S. Department of
  Commerce - The Smart Grid Interoperability Panel - Cyber Security Working
  Group}.
\newblock {NISTIR 7628 - Guidelines for Smart Grid Cyber Security: Vol. 3,
  Supportive Analyses and References}, August 2010.
\newblock
  \url{https://www.smartgrid.gov/document/nistir_7628_guidelines_smart_grid_cyber_security_vol_3_supportive_analyses_and_references}
  [Accessed: April 2020].

\bibitem{Sandia_MG}
Cynthia~K Veitch, Jordan~M. Henry, Bryan~T. Richardson, and Derek~H. Hard.
\newblock {\em {SANDIA REPORT: SAND2013-5472 - Microgrid Cyber Security
  Reference Architecture (Version 1.0)}}.
\newblock Sandia National Laboratories, 2013.

\bibitem{MITRE_ICS}
{MITRE}.
\newblock {ATT\&CK For Industrial Control Systems}.
\newblock \url{https://collaborate.mitre.org/attackics/index.php/Main_Page}
  [Accessed: April 2020].

\bibitem{MITRE_PRESS}
{MITRE}.
\newblock {Press Release: MITRE Releases Framework for Cyber Attacks on
  Industrial Control Systems}.
\newblock
  \url{https://www.mitre.org/news/press-releases/mitre-releases-framework-for-cyber-attacks-on-industrial-control-systems}
  [Accessed: April 2020].

\bibitem{7152015}
X.~{Zhong}, L.~{Yu}, R.~{Brooks}, and G.~K. {Venayagamoorthy}.
\newblock Cyber security in smart dc microgrid operations.
\newblock In {\em 2015 IEEE First International Conference on DC Microgrids
  (ICDCM)}, pages 86--91, 2015.

\bibitem{8953044}
N.~{Nikmehr} and S.~{Moradi Moghadam}.
\newblock Game-theoretic cybersecurity analysis for false data injection attack
  on networked microgrids.
\newblock {\em IET Cyber-Physical Systems: Theory Applications}, 4(4):365--373,
  2019.

\bibitem{7301476}
A.~{Teixeira}, K.~{Paridari}, H.~{Sandberg}, and K.~H. {Johansson}.
\newblock Voltage control for interconnected microgrids under adversarial
  actions.
\newblock In {\em 2015 IEEE 20th Conference on Emerging Technologies Factory
  Automation (ETFA)}, pages 1--8, 2015.

\bibitem{chlela_attack}
Martine Chlela, Geza Joos, and Marthe Kassouf.
\newblock Impact of cyber-attacks on islanded microgrid operation.
\newblock In {\em Proceedings of the Workshop on Communications, Computation
  and Control for Resilient Smart Energy Systems}, RSES ’16, New York, NY,
  USA, 2016. Association for Computing Machinery.

\bibitem{8550549}
A.~J. {Gallo}, M.~S. {Turan}, P.~{Nahata}, F.~{Boem}, T.~{Parisini}, and
  G.~{Ferrari-Trecate}.
\newblock Distributed cyber-attack detection in the secondary control of dc
  microgrids.
\newblock In {\em 2018 European Control Conference (ECC)}, pages 344--349,
  2018.

\bibitem{8051297}
M.~M. {Rana}, L.~{Li}, and S.~W. {Su}.
\newblock Cyber attack protection and control of microgrids.
\newblock {\em IEEE/CAA Journal of Automatica Sinica}, 5(2):602--609, 2018.

\bibitem{7829372}
O.~A. {Beg}, T.~T. {Johnson}, and A.~{Davoudi}.
\newblock Detection of false-data injection attacks in cyber-physical dc
  microgrids.
\newblock {\em IEEE Transactions on Industrial Informatics}, 13(5):2693--2703,
  2017.

\bibitem{HOSSAIN2014132}
Eklas Hossain, Ersan Kabalci, Ramazan Bayindir, and Ronald Perez.
\newblock Microgrid testbeds around the world: State of art.
\newblock {\em Energy Conversion and Management}, 86:132 -- 153, 2014.

\bibitem{6392520}
S.~{Glover}, J.~{Neely}, A.~{Lentine}, J.~{Finn}, F.~{White}, P.~{Foster},
  O.~{Wasynczuk}, S.~{Pekarek}, and B.~{Loop}.
\newblock Secure scalable microgrid test bed at sandia national laboratories.
\newblock In {\em 2012 IEEE International Conference on Cyber Technology in
  Automation, Control, and Intelligent Systems (CYBER)}, pages 23--27, 2012.

\bibitem{goarticle}
D.~Silver, J.~Schrittwieser, K.~Simonyan, and et~al.
\newblock Mastering the game of go without human knowledge.
\newblock {\em Nature}, 550(6):354--359, 2017.

\bibitem{starcraft}
{Deepmind}.
\newblock {AlphaStar: Mastering the Real-Time Strategy Game StarCraft II}.
\newblock
  \url{https://deepmind.com/blog/article/alphastar-mastering-real-time-strategy-game-starcraft-ii}
  [Accessed: April 2020].

\bibitem{sutton}
Richard~S Sutton and Adrew~G. Barto.
\newblock {\em {Reinforcement Learning: An Introduction}}.
\newblock A Bradford Book, 2nd edition, 2018.

\bibitem{8611595}
M.~C. {Ghanem} and T.~M. {Chen}.
\newblock Reinforcement learning for intelligent penetration testing.
\newblock In {\em 2018 Second World Conference on Smart Trends in Systems,
  Security and Sustainability (WorldS4)}, pages 185--192, 2018.

\bibitem{pomdp_hackers}
Carlos Saurraute, Olivier Buffet, and Jörg Hoffmann.
\newblock {POMDPs Make Better Hakers: Accounting for Uncertainty in Penetration
  Testing}.
\newblock In {\em 26th AAAI Conference on Artificial Intelligence (AAAI12)},
  pages 1816--1824, 2012.

\bibitem{mitre_mdp}
Andy Applebaum, Doug Miller, Blake Strom, Chris Korban, and Ross Wolf.
\newblock Intelligent, automated red team emulation.
\newblock In {\em Proceedings of the 32nd Annual Conference on Computer
  Security Applications}, ACSAC ’16, page 363–373, New York, NY, USA, 2016.
  Association for Computing Machinery.

\bibitem{MATLAB}
{MATLAB: MathWorks Documentation}.
\newblock {Simplified Model of a Small Scale Micro-Grid}.
\newblock
  \url{https://www.mathworks.com/help/physmod/sps/examples/simplified-model-of-a-small-scale-micro-grid.html}
  [Accessed: April 2020].

\bibitem{CPLEX_REF}
{IBM}.
\newblock {IBM ILOG CPLEX Optimization Studio}.
\newblock \url{https://www.ibm.com/products/ilog-cplex-optimization-studio}
  [Accessed: April 2020].

\bibitem{jeroslow}
Robert~G. Jeroslow.
\newblock {The polynomial hierarchy and a simple model for competitive
  analysis}.
\newblock {\em Mathematical Programming}, 32:146--164, 1985.

\bibitem{uciml}
{University of California, Irvine Machine Learning Repository}.
\newblock {Individual household electric power consumption Data Set}.
\newblock \url{https://archive.ics.uci.edu/ml/datasets/individual+household+
  electric+power+consumption} [Accessed: April 2020].

\bibitem{weatherstats}
{Canada Weather Stats}.
\newblock {Data Download for Ottawa (Kanata - Orléans)}.
\newblock \url{https://ottawa.weatherstats.ca/download.html} [Accessed: April
  2020].

\bibitem{ontariohydro}
{Ontario Hydro}.
\newblock {Ontario Hydro Rates}.
\newblock \url{http://www.ontario-hydro.com/current-rates} [Accessed: April
  2020].

\bibitem{ac_algo}
A.~{Steinbach}.
\newblock {Actor-critic using deep-RL: continuous mountain car in TensorFlow}.
\newblock
  \url{https://medium.com/@asteinbach/actor-critic-using-deep-rl-continuous-mountain-car-in-tensorflow-4c1fb2110f7c}
  [Accessed: April 2020].

\bibitem{sutton_policy}
R.~S. Sutton, D.~McAllester, S.~Sing, and Y.~Mansour.
\newblock Policy gradient methods for reinforcement learning with function
  approximation.
\newblock In {\em Advances in neural information processing systems (NIPS)},
  pages 1057--1063, 1999.

\bibitem{6392457}
I.~{Grondman}, L.~{Busoniu}, G.~A.~D. {Lopes}, and R.~{Baduska}.
\newblock A survey of actor-critic reinforcement learning: Standard and natural
  policy gradients.
\newblock {\em {IEEE Transactions on Systems, Man, and Cybernetics, Part C
  (Applications and Reviews)}}, 42(6):1291--1307, 2012.

\end{thebibliography}
\end{document}